\documentstyle[12pt]{article}
\begin{document}

\begin{titlepage}
\baselineskip .15in
\begin{flushright}
WU-AP/40/93
\end{flushright}

{}~\\

\vskip 1.5cm
\begin{center}
{\bf
\vskip 1.5cm
{\large\bf Non-Abelian Black Holes and
Catastrophe Theory II:\\[.5em]
Charged Type}

}\vskip .8in

{\sc T. Tachizawa}$^{(a)}$, {\sc K. Maeda}$^{(b)}$ and
{\sc T. Torii}$^{(c)}$  \\[.5em]
{\em Department of Physics, Waseda University,
 Tokyo 169, Japan}\\[1em]

\end{center}
\vfill

\begin{abstract}

We reanalyze the gravitating monopole and its black hole
solutions in the Einstein-Yang-Mills-Higgs system and we
discuss their stabilities  from the point of view of
catastrophe theory.  Although these non-trivial  solutions
 exhibit fine and  complicated structures,
we find that stability is systematically
understood via a swallow tail
catastrophe.
The Reissner-Nordstr\"{o}m trivial solution becomes
unstable from the point where the non-trivial monopole
black hole appears.  We also  find that, within a very small
parameter range, the specific heat of a monopole black hole
changes its sign .

\end{abstract}


\vfill
\begin{center}
September, 1994
\end{center}
\vfill
(a)~~electronic mail:63L507@cfi.waseda.ac.jp \ \ \
                         JSPS Research Fellow \\
(b)~~electronic mail:
maeda@cfi.waseda.ac.jp\\  (c)~~electronic mail:
64L514@cfi.waseda.ac.jp

\end{titlepage}

\normalsize
\baselineskip = 22pt


\section{Introduction}

The magnetic monopole was first  discussed by Dirac in the
context of the $U(1)$ gauge field\cite{Dirac}.
\mbox{'t Hooft} and Polyakov found a new type of
monopole solution in the $SO(3)$
Yang-Mills-Higgs system\cite{'tHooft-Polyakov}. Later on we
realized that  such a type of  monopole
usually  appears during a cosmological phase transition
of gauge symmetries based on the grand unified
theories (GUTs).
Since it is a non-trivial structure in the
Yang-Mills-Higgs system, there may be many interesting
points to be investigated.
Among them, from the viewpoint of cosmology, we may
address the following questions: Does it really exist and
is it stable?   How was it created in the history of the
universe?  What are the cosmological implications?
Can we observe it now?

One important outcome of such research is the
so-called inflationary scenario in the early universe.
Because the thermal production of such monopoles  in the
early universe results in destruction in the conventional
Big Bang scenario,  the idea of inflation was proposed in
the early 80's to resolve this monopole problem.  As a
result, we may not find so many GUTs-monopoles in the
present epoch. Although the present amount of such a massive
monopole is very much constrained within the present
observation as well as by the inflationary scenario, there
may have existed some unknown important  effect in the early
stage of the universe.  For example,  Vilenkin and Linde
have recently proposed a new type of inflationary scenario,
which occurs inside of topological defects such as
monopoles at the Planck energy scale\cite{top-inf}.

Since a monopole in the GUTs or at the Planck energy scale
is very massive, a gravitational effect may become
important.  Breitenlohner  et al.\cite{BFM},
Lee et al.\cite{LNW}, and Ortiz\cite{Ortiz}
recently found a gravitating monopole and its black hole
solutions (monopole black hole).    Although the role of
such objects  in the history of the universe is not yet
clear, we have found several  important results from a
fundamental viewpoint.   For example, because of the
existence of this non-trivial structure, the trivial
charged black hole  (Reissner-Nordstr\"{o}m solution)
becomes unstable. However, as pointed out by Aichelburg and
Bizon\cite{AB}, mass energy may not be a good indicator
of stability when more than two non-trivial solutions
exist.   In order to find a universal picture of such
non-trivial structures and investigate the role of those
objects in the early universe, we need more detailed
investigations.  This is the purpose of the
present paper.

As for non-trivial gravitating structures including
black hole solutions, we know, so far, several
numerical solutions\cite{TMT} (Paper I. see also references
therein).  Because neither the vacuum Einstein nor the pure
Yang-Mills system has a  non-singular finite energy
solution, the first discovery of a  particle-like solution
in the Einstein-Yang-Mills system  by Bartnik and
Mckinnon\cite{BM} was surprising.  After this discovery, the
so-called colored black hole  was also  discovered in the
same system\cite{CBH}.  This colored  black hole has a
non-Abelian hair, which does not contribute to a global
charge.  Thus this casts doubt on the no-hair
conjecture about black holes.  These examples suggest that a
non-Abelian(NA) field coupled to gravity throws a new light
upon general relativity.  The monopole and its black holes
also provide us with a candidate for a counterexample to
the no-hair conjecture. The trivial Reissner-Nordstr\"{o}m
black hole becomes unstable when those non-trivial
structures  exist and then goes to the stable monopole black
hole. In this sense, the monopole  black hole could be one
of the most plausible counterexamples.   To discuss the
evolution of those structures, we have to know their
properties and  stabilities. That is another point to
be discussed here.    Apart from the monopole and its black
holes, we presented a universal picture of NA black holes in
\cite{MTTM,TMT} from the point of view of catastrophe
theory. In those papers, we analyzed  mainly globally
neutral black holes.  In particular, we found  that the
stability of the black holes was naturally  understood by
catastrophe theory.

In this report, we will focus on the charged case,
that is,  the Einstein-Yang-Mills system with  the  real
triplet Higgs field.  A particle-like solution in this
system is a gravitating 't Hooft-Polyakov magnetic
monopole.     Hence, the black hole in this system is
globally charged.    The properties of the globally charged
hole are very different from those of the neutral cases, as
we shall see later. Nonetheless,  we will show that  we can
again apply  catastrophe theory.  Thus, catastrophe
theory not only is a powerful tool to treat the stability
problem  but also gives us a universal picture of NA black
holes.

The plan of this paper is as follows.    In Sec.~II, we
derive the basic equations.   In Sec.~III, we briefly review
studies of the gravitating monopole and its black holes and
summarize the known results.    In Sec.~IV, we discuss the
stability of the non-trivial structures via catastrophe
theory.    In Sec.~V, we discuss a new feature of the
monopole black hole obtained from the present analysis
and also show its spacetime structure.   In
Sec.~VI, we give our conclusion and some final remarks.


\vskip 2 cm
\setcounter{equation}{0}
\section{BASIC EQUATIONS}

The $SO(3)$ EYMH system is described by the action:
\begin{equation}
  S=\int d^4 x \sqrt{-g} \left[\frac{1}{16\pi G}R
     +{\cal L}_{matter}\right], \label{action}
\end{equation}
with
\begin{equation}
  {\cal L}_{matter}=-\frac{1}{4}F^a_{\mu\nu}F^{a\mu\nu}
                    -\frac{1}{2}D_{\mu}\Phi^a D^{\mu}\Phi^a
                    -\frac{\lambda}{4}\left( \Phi^a \Phi^a
                    -v^2 \right)^2,
\end{equation}
\begin{eqnarray}
F^a_{\mu\nu} & = & \partial_{\mu}A^a_{\nu}
  -\partial_{\nu}A^a_{\mu}
  -e\epsilon^{abc}A^b_{\mu}A^c_{\nu},
\\
  D_{\mu}\Phi^a & = & \partial_{\mu}\Phi^a
            +e\epsilon^{abc} A^b_{\mu}\Phi^c,
\end{eqnarray}
where $A^a_{\mu}$ and $F^a_{\mu\nu}$ are the $SO(3)$
Yang-Mills field potential and its field strength,
respectively, and
$\Phi^a$ is the real triplet Higgs field.
$D_{\mu}$ is the covariant
derivative.    $G$, $e$, $v$ and $\lambda$ are the
Newton's gravitational constant, the gauge coupling
constant, the vacuum expectation value of the Higgs field,
and the Higgs self coupling constant, respectively.     We
use the unit
$c=\hbar=1$.      $M_{Pl} \equiv 1/\sqrt{G}$ is the Planck
mass.

We consider a static and spherically symmetric solution.
The metric is written as
\begin{equation}
  ds^2=-\left(1-\frac{2Gm(r)}{r}\right)e^{-2\delta(r)}dt^2
       +\left(1-\frac{2Gm(r)}{r}\right)^{-1}dr^2
       +r^2(d\theta^2+\sin^2\theta d\phi^2).
\label{metric}
\end{equation}
For the matter fields, we adopt the so-called hedgehog
ansatz given by
\begin{eqnarray}
  \Phi^a & = & v \hat{r}^a h(r),
\\
  A^a_i & = & \omega_i^c
\epsilon^{cab}\hat{r}^b\frac{1-w(r)}{er}, \ A^a_0=0,
\end{eqnarray}
where $\hat{r}^a$ and $\omega_i^c$ are a unit radial vector
in the internal space and  a triad, respectively.
Introducing dimensionless variables,
\begin{equation}
  \tilde{r}=evr, ~~~\mbox{and}~~~  \
\tilde{m}=evm/M_{Pl}^2,
\end{equation}
and dimensionless parameters,
\begin{equation}
  \tilde{v}=v/M_{Pl},  ~~~\mbox{and}~~~
\ \ \tilde{\lambda}=\lambda/e^2,
\end{equation}
the Einstein equations and the field equations derived from
the action (\ref{action})   are reduced to
\begin{eqnarray}
  \frac{d\delta}{d\tilde{r}} & = & -8\pi \tilde{v}^2
\tilde{r}\left[
  \frac{1}{\tilde{r}^2}\left( \frac{dw}{d\tilde{r}}
  \right)^2+\frac{1}{2}\left(
\frac{dh}{d\tilde{r}}\right)^2
\right],  \label{Ein1}\\
  \frac{d\tilde{m}}{d\tilde{r}}& = &4\pi
  \tilde{v}^2 \left[
\left(1-
  \frac{2 \tilde{m}}{\tilde{r}}\right) \left\{ \left(
\frac{dw}{d\tilde{r}}
  \right)^2+\frac{\tilde{r}^2}{2} \left(
\frac{dh}{d\tilde{r}} \right)^2
  \right\}+\frac{(1-w^2)^2}{2\tilde{r}^2}+w^2 h^2 \right]
\nonumber
\\
  & & ~~~ +\pi\tilde{\lambda}\tilde{v}^2 \tilde{r}^2
(h^2-1)^2
\
  (=-\frac{4\pi G}{e^2 v^2} {T^0}_0 \tilde{r}^2),
\label{Ein2}
\\[.3em]
  \frac{d^2 w}{d\tilde{r}^2}&=&\frac{1}{\tilde{r}^2\left(
  1-2\tilde{m}/\tilde{r}
  \right)}\left[
  -w(1-w^2-\tilde{r}^2 h^2)
  -2\tilde{m}\frac{dw}{d\tilde{r}}\right.
  \nonumber \\
  & &  ~~~ \left.+8\pi\tilde{v}^2
  \tilde{r}\frac{dw}{d\tilde{r}}\left\{
  \frac{(1-w^2)^2}{2\tilde{r}^2}+w^2 h^2 +\tilde{\lambda}
  \frac{\tilde{r}^2(h^2-1)^2}{4}\right\}\right],
\label{YM}
\\[.3em]
  \frac{d^2 h}{d\tilde{r}^2}&=&
  -\frac{2}{\tilde{r}}\frac{dh}{d\tilde{r}}+
  \frac{1}{\tilde{r}^2\left(1-2\tilde{m}/\tilde{r}
  \right)}\left[ 2hw^2+\tilde{\lambda} \tilde{r}^2 h(h^2-1)
  -2\tilde{m}\frac{dh}{d\tilde{r}}\right.\nonumber \\
  & &  ~~~ \left.+8\pi\tilde{v}^2
  \tilde{r}\frac{dh}{d\tilde{r}}\left\{
  \frac{(1-w^2)^2}{2\tilde{r}^2}+w^2 h^2 +\tilde{\lambda}
  \frac{\tilde{r}^2(h^2-1)^2}{4}\right\}\right],
\label{Higgs}
\end{eqnarray}
We choose the normalization of $t$ by setting
$\delta(\infty)=0$.

The boundary conditions of a globally regular monopole
solution are
\begin{equation}
  w(\infty)=0, \ \  h(\infty)=1, \ \
             \tilde{m}(\infty) = {\rm finite},
\label{bc_reg1}
\end{equation}
at
spatial infinity and
\begin{eqnarray}
  w(\tilde{r}) & = & 1-c_w \tilde{r}^2 + \cdots ,
\label{bc_reg2}\\
  h(\tilde{r}) & = & c_h \tilde{r} + \cdots ,
\label{bc_reg3}\\
  \tilde{m}(\tilde{r}) & = & \frac{4\pi}{3}\tilde{v}^2
\left(6{c_w}^2
+\frac{3}{2}{c_h}^2+\frac{1}{4}\tilde{\lambda}
  \right)\tilde{r}^3+ \cdots , \label{bc_reg4}
\end{eqnarray}
near the origin,
where $c_w$ and $c_h$ are constants
 determined iteratively  so that $w$
and $h$ satisfy the boundary condition (\ref{bc_reg1}).

As for the black hole solution, the boundary condition at
spatial infinity is the same as that of the regular
monopole  solution while that at the horizon is  given by
\begin{equation}
  \tilde{m}(\tilde{r}_H)=\frac{\tilde{r}_H}{2},
\end{equation}
where $\tilde{r}_H$ is the value of $\tilde{r}$ at the
horizon. Moreover the
square brackets in (\ref{YM}) and (\ref{Higgs}) must
vanish at $\tilde{r}_H$ for the horizon to be regular.
Hence,
\begin{equation}
  \left. \frac{dw}{d\tilde{r}}
\right|_{\tilde{r}=\tilde{r}_H}
  =\frac{w_H(1-w_H^2-h_H^2\tilde{r}_H^2)}
  {2\pi\tilde{v}^2 \tilde{r}_H
  \left[2\tilde{r}_H^{-2}(1-w_H^2)^2+ 4w_H^2 h_H^2 +
  \tilde{\lambda}\tilde{r}_H^2(h_H^2-1)^2 \right]
-\tilde{r}_H},
\end{equation}
\begin{equation}
  \left. \frac{dh}{d\tilde{r}} \right|_{\tilde{r}
=\tilde{r}_H} =\frac{-h_H
  [2w_H^2+\tilde{\lambda} \tilde{r}_H^2
 (h_H^2-1)]}
  {2\pi\tilde{v}^2 \tilde{r}_H
  \left[2\tilde{r}_H^{-2}(1-w_H^2)^2+ 4w_H^2 h_H^2 +
  \tilde{\lambda}\tilde{r}_H^2(h_H^2-1)^2 \right]
-\tilde{r}_H},
\end{equation}
where
\mbox{$w_H\equiv w(\tilde{r}_H)$} and \mbox{$h_H\equiv
h(\tilde{r}_H)$}.
$w_H$ and $h_H$ are shooting parameters and should be
determined iteratively so that the boundary condition
(\ref{bc_reg1}) is satisfied. Therefore, this is  a kind of
eigenvalue problem though the equations do not contain a
physical constant parameter which becomes an eigenvalue.


\vskip 2 cm
\setcounter{equation}{0}
\section{OVERVIEW OF A MONOPOLE BLACK HOLE}

So far, several research groups have investigated the
gravitating monopole and its black hole solutions and
found very fine and  complicated  structures in a solution
space. Here we shall briefly review those non-trivial
structures and summarize their properties.

\noindent
{\bf (1)} Monopole:\\
Breitenlohner  et al.\cite{BFM}, Lee et al.\cite{LNW},
and Ortiz\cite{Ortiz} solved Eqs.(\ref{Ein1}), (\ref{Ein2})
numerically under the boundary conditions
(\ref{bc_reg1})-(\ref{bc_reg4}), and found a gravitating
monopole.     There exists a maximal value of $\tilde{v}$,
$\tilde{v}_{\rm max} (\sim 0.3958)$\cite{notation}  for
$\tilde{\lambda}=0$, beyond which no solution
exists\cite{BFM}.   Only a trivial RN black hole solution
can exist.   We  understand the existence of this critical
value of $\tilde{v}$ intuitively as follows:  The mass of
monopole and its core radius are $\sim 4\pi v/e$ and  $\sim
1/ev$, respectively.  Then, as $v$ gets large, the monopole
radius decreases while  its gravitational radius
(\mbox{$8\pi G  v/e$}) increases.  The monopole radius
eventually becomes smaller than its gravitational  radius
and  it collapses into a black hole.  This happens when
\mbox{$v\sim M_{Pl}/\sqrt{8\pi}$}. For $e\ll 1$, since
the energy \mbox{density$\sim e^2 v^4
\ll {M_{Pl}}^4$}, then we can still ignore the effect of
quantum gravity.

Near the critical point $\tilde{v}_{\rm max}$, there exist
two monopole solutions with different mass $M$ at least for
small values of  $\tilde{\lambda}$ (see Fig.~1).
One is more massive than the other and it disappears at
some  value of $\tilde{v}$,
$\tilde{v}_{\rm extreme} (<\tilde{v}_{\rm max})$, where
the  solution turns out to be the extreme black hole
solution. The more massive branch is unstable while the
less  massive branch is stable\cite{Hollmann}.  Since a
cusp appears at $\tilde{v}_{\rm max}$ (see Fig.~1), we can
understand this stability via catastrophe theory  (see
later).

There are two interesting limiting cases: (i) $G=0$ and
(ii) $v=0$.     $\tilde{v}=0$ denotes $G=0$, i.e., gravity
is switched off.   (Notice that the limit of $\tilde{v}=0$
does not  recover $v=0$  because $\tilde{r}$ contains
$v$.)   We find the solutions in a flat spacetime. In the
limit of $\tilde{\lambda}=0$ with  $\tilde{v}=0$ (i.e., flat
spacetime), namely, in the Bogomol'nyi-Prasad-Sommerfield
limit\cite{Prasad_Sommerfield}, we have the following
analytic solution:
\begin{equation}
  w(\tilde{r})=\frac{\tilde{r}}{\sinh \tilde{r}}\ ,
\ \ \ h(\tilde{r})=\coth \tilde{r}-\frac{1}{\tilde{r}}.
\end{equation}

On the other hand, in the limit of $v\rightarrow 0$ keeping
$G$ fixed, there is no solution
which satisfies the above boundary conditions.   Instead,
if we set $h\equiv 0$, the solution is nolonger a monopole
solution but the Bartnik-McKinnon(BM) solution in the
Einstein-Yang-Mills system\cite{BM}.  The typical mass
scale of the BM particle is $\sqrt{4\pi}M_{Pl}/e$, which is
the the same as that  of the extreme Reissner-Nordstr\"{o}m
black hole with a magnetic charge
$1/e$.

\noindent
{\bf (2)} Black holes:\\
The Reissner-Nordstr\"{o}m(RN) solution always exists for
arbitrary  values of $\tilde{v}$ and $\tilde{\lambda}$.
It is a trivial solution in this system, i.e.,
\begin{equation}
 w(r)\equiv 0,  \ \  h(r)\equiv 1, \ \
 m(r)=M-\frac{2\pi}{e^2 r}, \ \  ~~{\rm and}~~
\delta(r)\equiv 0,
  \ \
\end{equation}
where $M$ is the gravitational mass.  On the other hand, a
non-trivial solution exists only for a finite  range of
$\tilde{v}$ just as in the case of a monopole. We may
regard the NA black hole as a black hole lying inside a
't Hooft-Polyakov monopole.    Therefore, we shall call
the NA black hole a monopole black hole.

Breitenlohner et al.\cite{BFM} solved the above equations
(\ref{Ein1})-(\ref{Higgs}) numerically and showed
explicitly the monopole black hole solution.   Lee et
al.\cite{LNW} suggested the existence of the monopole black
hole solution from consideration of a simplified model.
They assumed that the distribution of the energy density is
\begin{equation}
  \rho=\left\{ \begin{array}{ll}
            \rho_0 \ (\mbox{const.}) & (r<R) \\
            1/2e^2 r^4     & (r>R)
               \end{array} \right. \label{simplifyed}
\end{equation}
where $R$ denotes the core radius of a monopole.
The mass function is obtained by integration
\begin{equation}
  m(r)=\int_{0}^{r} 4\pi
  r^2 \rho(r)dr + m(0).
\end{equation}
(Later we will also discuss the validity of this model and
use it to show the Penrose diagram.)

Roughly speaking, the NA black hole can exist when the
horizon radius ($\sim GM$) is less than the monopole radius
($\sim 1/ev$), and it is stable.
This stability may be naively understood from the
following argument(but see \S.4 for the details): As the
horizon radius $r_H$ shrinks, the energy density of the
magnetic field near the horizon gets as large as
\mbox{$\sim 1/r_H^4$} (see eq.(\ref{den1})-(\ref{den6})).
Hence, the energy density inside the core is lowered and is
energetically favorable to restore symmetry and to shield
the magnetic field in the monopole core.     Such is the
case of the monopole black hole.

Depending on the parameters $\tilde{\lambda}$ and
$\tilde{v}$, a variety of solutions are found.   If
$\tilde{\lambda}$ is large, e.g. $\tilde{\lambda}=1$,  we
have a rather simple structure, that is, there is only one
NA  black hole branch in addition to the RN branch.   The
two branches merge at the bifurcation point
$B$.  However, if $\tilde{\lambda}$ is small enough, e.g.
$\tilde{\lambda}=0.1$, the NA solutions exist in two
branches and show a cusp $C$. The bifurcation point $B$
where the NA black hole and the RN branch merge also exist.
In a small range of parameters, there are three black hole
solutions.  Which one is stable?    Aichelburg and
Bizon\cite{AB} claimed that the mass is not a good
indicator of stability. Because of such a complexity
in the EYMH system, they also studied a simpler case of
$\tilde{\lambda}=\infty$, although it turns out that the
case of $\tilde{\lambda}=\infty$ is not the same as the case
of \mbox{$\tilde{\lambda}=$finite}.   Later we will show a
unified picture via catastrophe theory.

\noindent
{\bf (3)} Higher excited modes:\\
Besides  a fundamental solution, in which
$h$ and $w$ are monotonic functions of $r$, there exists a
discrete family of radial excitations with an increasing
number of zeros of $w$\cite{BFM}.  The inside structure of
the solution in this family approaches the BM solution
in the limit of  $v\rightarrow 0$, although the behaviors
at large distances are not the same because of different
boundary conditions.

\noindent
{\bf (4)} Stabilities:\\
One of the most important properties of those non-trivial
structures is stability.

The trivial RN black hole seems to be stable, but it turns
out to be  unstable against linear radial perturbation for
some range of parameters\cite{LNW}.
Then, the stable non-trivial monopole black hole appears.

When we have more than two solutions, the less massive one
seems to be more stable because it is energetically more
favorable.   The result in the present case, however, is
not so simple.    As discussed in (2), there are three
solutions (one RN and two monopole BHs) in some parameter
range.   The smallest mass black hole is of course stable,
but one of the heavier holes is also stable at least
against linear radial perturbations.   To understand this
situation, we shall adopt catastrophe theory and show in
the next section how stability is naturally understood.

\noindent
{\bf (5)} Thermodynamical properties:\\
A RN black hole changes the sign of its specific
heat when the charge becomes larger than the
critical value $Q_{cr}=\sqrt{3}M/2$.   Davies claimed that
it is a second order phase transition\cite{Davies}.   This
may indicate the stability change of the charged black hole
in the heat bath system at the critical value.   How about
in the EYMH theory?  \   Lee et al. claimed that the
specific heat of the NA bole is always negative\cite{LNW},
but as we discuss in \S.5, there is a very narrow range of
parameters where the specific heat becomes positive.

\noindent
{\bf (6)} Fate of charged black hole:\\
What  happens on such a small black hole when we take into
account the Hawking  evaporation process?    In the
Einstein-Maxwell theory, as the RN black hole loses its
mass via the Hawking thermal radiation, if the charge is
conserved,  the black hole approaches the extreme one and
the temperature goes to zero. The evaporation will cease.
Hence, the extreme RN black hole is stable and it may
become a cosmological remnant.

On the other hand, in the Einstein-Yang-Mills case, the RN
black hole becomes classically unstable when its horizon
radius is less than the monopole radius
$\sim 1/ev$\cite{LNW}.    Thus, as the RN black hole
evaporates, it turns out to be  unstable and probably
becomes a monopole black hole.
Since the specific heat of the stable monopole black hole is
negative\cite{LNW} (but see later for the unstable one),
the monopole black hole does not stop evaporating.   Then,
we will find a regular monopole at the end.


\vskip 2 cm
\setcounter{equation}{0}
\section{STABILITY OF A MONOPOLE BLACK HOLE
         AND CATASTROPHE THEORY}

When we dealt with neutral black holes\cite{MTTM,TMT},
the mass-horizon radius diagram played an important role.
The cusp structure appeared in the diagram, and it gave us
a clue to apply catastrophe theory.  Here, we again  begin
with similar diagrams. For a review of catastrophe
theory, see, for example, reference\cite{Poston}, and for
its application to physics, reference\cite{Kusmartsev}.

Fig.~2 shows the relation between the mass of black
holes and  their horizon radius when $\tilde{\lambda}=1$ and
$\tilde{v}=0.05$.   ($\tilde{M}=\tilde{m}(\infty)$)   The
mass of the RN  black hole has a lower bound (the extreme
case) because its  magnetic charge is fixed as $1/e$.  From
Fig.~2, unlike in the neutral case, we cannot find any cusp
structure.

On the other hand, when $\tilde{\lambda}=0.1$ and
$\tilde{v}=0.05$, the relation becomes as shown in
Fig.~3(a).  This looks similar to Fig.~2 at first
glance.  However, as shown in Fig.~3(b), which is an
enlarged diagram near the bifurcation point $B$, where the
branch of the NA solutions and that of the RN solutions
merge, a cusp structure appears.

So, we have two behaviors of the solutions depending on the
value of $\tilde{\lambda}$.   From our analysis of the
neutral case and the discussion by Lee et al.\cite{LNW} of
the stability of the RN black hole, we can guess the
stability of the solution as follows: In the first case, as
in Fig.~2, the NA solution is always stable. While, the RN
solution is stable when the mass is larger than $M_{\rm
N-R}$, which is the mass at the bifurcation point, $B$, but
it becomes unstable if it is less than $M_{\rm N-R}$.  In
the second case, as in Fig.~3, the NA solution is stable in
the branch of the larger horizon radius,
$AC$, and the stability changes at the cuspidal edge $C$.
It becomes unstable in the branch of the smaller horizon
radius, $CB$.  The RN solution is stable when the mass is
larger than $M_{\rm N-R}$, but otherwise it is unstable.
We will confirm this below via catastrophe theory.

Although we divide the solutions into two classes above,
since our discussion is based on the numerical results, we
cannot exclude the possibility that a cusp structure may
also exist in the case of Fig.~2 but is so small that we
cannot see it.    Catastrophe theory, however, explains
very naturally the existence of two classes  depending on
the parameters, and the change of the stability described
above.  Which elementary catastrophe explains these
behaviors?  We find that it is a swallow tail type.

In catastrophe theory, we discuss changes in the shape of
a potential function as we change control parameters.
Thom's theory guarantees that if a system which is
described by any potential function and has less than or
equal to four control parameters experiences catastrophic
change, the potential function can be made to coincide with
one of the seven elementary catastrophe's potential
functions by diffeomorphism in the neighborhood of the point
where the catastrophe occurs.  Of course, we assume in the
above theorem that the system is structural stable (see,
for example, reference\cite{Poston}).  One or two state
variables characterize the system, and the minimal and the
maximal points of the potential function represent a stable
and an unstable configuration,
respectively\cite{stable}.

Any potential function of elementary catastrophe has one
minimal and one maximal point at least.  Ordinary black
holes belonging to the Kerr-Newman family have only three
hairs.  If global charges(mass, angular momentum, and
electric or magnetic charge) are given, the
configuration is uniquely determined.  On the other hand,
NA black holes can violate the no-hair conjecture,
and given global charges do not necessarily determine the
configuration completely.  That is, even if control
parameters are given, several stationary points in the
potential function are possible.  This is why catastrophe
theory is applicable in analyzing the stability of NA black
holes.  As for the stability of black holes belonging
to the Kerr-Newman family, see reference\cite{Katz}.

A swallow tail catastrophe has three control parameters,
$a$,
$b$, $c$, and one state variable, $x$, and its potential
function is described as
\begin{equation}
 V=\frac{1}{5}x^5+\frac{1}{3}ax^3+\frac{1}{2}bx^2+cx.
\end{equation}
A set of the points where catastrophe occurs, that
is, a catastrophe set, is schematically shown in Fig.~4,
which looks like a swallow's tail.
In our present model, we set $c=0$ and reduce the
three-dimensional control parameter space to a
two-dimensional one.  We need this  reduction in order that
the RN solution ($x=0$) can always exist as a trivial
solution.   A control parameter $a$ depends on
$\tilde{\lambda}$ and
$\tilde{v}$, and $b$ depends on   $\tilde{M}
\mbox{(}\equiv ev M/M^2_{Pl}\mbox{)}$ and $\tilde{v}$ as
\begin{equation}
  a=a(\tilde{\lambda}, \tilde{v}), \ \ b=b(\tilde{M},
\tilde{v}).
\end{equation}
As we shall soon see below, the condition of $a=0$
divides the solutions into the two cases of whether the
cusp exists or not.   That is to say, a cusp structure
appears in the tail of the swallow, while it does not
appear in its body.

We denote the value of $\tilde{\lambda}$ corresponding to
the case of $a=0$ as  $\tilde{\lambda}_{crit}$.
$\tilde{\lambda}_{crit}$ is a function of $\tilde{v}$.
Fig.~5 shows the schematic potential in the case of
$a>0$, namely,
$\tilde{\lambda}> \tilde{\lambda}_{crit}$.\
We regard the entropy of the black hole ($S=\mbox{area}/4$)
as the potential function.    When we consider the
entropy, the maximal point represents the stable
configuration.   Hence, we put $S=-V$.  The ordinate
represents the entropy $S$.   The abscissa
represents a state variable.  We adopt
$\delta_H$, which is the value of $\delta(r)$ at the
horizon, as the state variable.  We may adopt any
variable as a state variable as long as it can be mapped
to $\delta_H$ by diffeomorphism.  From our numerical
calculation, it seems that the ``effective magnetic
charge" of the black hole defined by
\begin{equation}
(1/4\pi)\int_{r=r_H} F = (1-{w_H}^2)/e ,
\end{equation}
or the minimum value of the function
$(1-2\tilde{m}/\tilde{r})$ is related to $\delta_H$
by diffeomorphism once we fix $\tilde{v}$ and
$\tilde{\lambda}$. Here, we take
$\delta_H$ because we can also use it when we discuss the
non-singular monopole.

Furthermore, we have to cut down the region of
$\delta_H<0$  because such a region  is unphysical.  We can
easily show that $\delta_H$ is always positive from
(\ref{Ein1}). (Remember that
$\delta(\infty)=0$.)  The maximal points in Fig.~5
represent stable solutions and the minimal points
unstable ones.  The origin ($\delta_H=0$) is always
a stationary point by virtue of setting $c\equiv 0$, and
it corresponds to the RN solution.

To discuss dynamical stability, we shall  consider a
physical process increasing the mass of the black hole.
Physically, it can be realized by matter accretion. We fix
parameters in the theory, i.e., $\tilde{v}$ and
$\tilde{\lambda}$.  Only $b$ varies by a change of the
mass.  We assume that
$b(\tilde{M},\tilde{v})$ is a monotonically increasing
function of $\tilde{M}$.  When $b$ is negative, the shape
of the potential function is shown in \mbox{Fig.5(a)}.  At
this time, the RN solution is unstable and the NA  solution
is stable.    Since our potential function is entropy, the
maximal point represents a stable solution and the minimal
one an unstable one.     Only the NA black hole can
really exist. As we increase $b$, the minimal value
increases and the minimal point approaches the
origin(\mbox{Fig.~5(b)}).  When
$b=0$, the minimal and the maximal points eventually
merge (Fig.5(c)).  This is the bifurcation point $B$ in
Fig.~2.    At this point, the NA black hole turns into
the RN black hole continuously.  Then the NA solution
disappears, and the RN solution begins to be
stable(Fig.5(d)).

 We can discuss the inverse process
decreasing the mass through the Hawking radiation.
Suppose there exists a
RN black hole, which is heavier than the mass at the
bifurcation point $B$.
When  its mass decreases,  the RN black hole
eventually reaches  the bifurcation point
$B$ and it changes to the RN black hole continuously.
It continues to evaporate, then probably turns into a
non-singular monopole\cite{LNW}.

Next we consider the case of
$a<0$ ($\tilde{\lambda} <\tilde{\lambda}_{crit}$).  In this
case, a cusp structure exists.   We again discuss the
case increasing $b$ from a negative  value.
The change of the potential form is shown in  Fig.~6.
Initially, one unstable RN solution ($R$) and one stable
NA solution ($N_1$)  exist.  Then, only the NA black hole
can really exist(Fig.6(a)).  As $b$ increases,
another pair of a maximal and minimal point ($N_2,
S$) comes to appear in the  unphysical region (Fig.6(b)).
Then the minimal point $N_2$ merges with the
point $R$ at $b=0$  (Fig.6(c)),
and the RN solution becomes stable.      However, since the
NA solution $N_1$ is still stable, nothing  happens at
this point.  And another branch of unstable NA
solutions ($N_2$) bifurcates from the branch of the RN
solution (Fig.6(d)).   Then, two stable solutions ($R,
N_1$) and one unstable solution ($N_2$) exist at the
same time.  That is, even if the global charge (=mass) is
fixed, two stable solutions become possible.  This means
a violation of the weak no-hair
conjecture\cite{AB,grqc2016}.  As $b$ increases further,
the stable and  unstable NA solutions approach
each other, and eventually they merge to become an
inflection point (Fig.6(e)).  This causes catastrophe. The
NA black hole jumps to a RN black hole discontinuously as
a solid arrow.   After that only the stable RN solution
remains (Fig.6(f)).

The inverse process, i.e., the case decreasing
$b(\tilde{M},\tilde{v})$, is described as follows:  Suppose
there exists only a stable RN solution ($R$)  initially.
As $b$ decreases, a pair of stable and unstable NA
solutions ($N_1, N_2$) appears (Fig.6(e)).  At this point,
since the RN solution remains  stable, nothing happens.
When $b$  decreases further and reaches zero, the stable RN
($R$) and the unstable NA branches ($N_2$) merge.   Then,
the RN solution begins to be unstable (Fig.6(c)).  The RN
black hole ($R$)  jumps to the NA black  hole ($N_1$)
discontinuously as the dotted arrow (Fig.6(b)).  Only the
NA solution remains as a stable solution.    Then, the NA
black hole will evaporate to be a stable monopole in the
same way as the case of
$\tilde{\lambda}>\tilde{\lambda}_{crit}$.

As we have seen above, a swallow tail
catastrophe can explain the behavior  of solutions.
Moreover, note that the RN branch  and the NA branch
always merge at $b=0$ (i.e., $\tilde{M}=\tilde{M}_{\rm
N-R}$) independently of the value of $\tilde{\lambda}$.  We
can confirm the fact that $\tilde{M}_{\rm N-R}$ is
independent of the value of $\tilde{\lambda}$ from
Fig.~7, which shows the critical values of mass at the
bifurcation points.   If
$\tilde{\lambda}<\tilde{\lambda}_{crit}$,
there are two critical values,
$\tilde{M}_{\rm
N-R}$ and $\tilde{M}_{\rm N-N}$, which correspond to the
bifurcations of $R$ and $N_2$ (Fig.6(c)) and of $N_1$ and
$N_2$ (Fig.6(e)) , respectively.
On the other hand, if
$\tilde{\lambda}>\tilde{\lambda}_{crit}$,
there is only one bifurcation of $R$ and $N$
(Fig.5(c))  appearing at $\tilde{M}_{\rm N-R}$.
{}From Fig.~7, we find that  $\tilde{M}_{\rm N-R}$
is in fact
independent  of
$\tilde{\lambda}$.

The dependence of $\tilde{M}_{\rm N-R}$ on $\tilde{v}$ is
shown in Fig.~8.   This relation is approximated
 as
\begin{equation}
  \tilde{M}_{\rm N-R}=7.256 \tilde{v}^{1.745}+0.273.
                           \label{fitting}
\end{equation}

We can  compare this with  the result of Aichelburg and
Bizon\cite{AB} who dealt with the case of
$\tilde{\lambda}=\infty$.  According to their result,
at
$\tilde{v}=0.288/\sqrt{4\pi}\approx 0.0812$,
$\tilde{M}_{\rm N-R}=1.25\times 0.288\approx 0.360$.
{}From our analysis (\ref{fitting}), we can extrapolate and
guess the value of $\tilde{M}_{\rm N-R}$ at
$\tilde{\lambda}$ as
\begin{equation}
  \tilde{M}_{\rm N-R}=7.256\times \left( 0.288/\sqrt{4\pi}
  \right)^{1.745}+0.273\approx 0.363.
\end{equation}
Those two values of $\tilde{M}_{\rm N-R}$ agree well. Thus,
we find that $\tilde{M}_{\rm N-R}$ is independent of the
 $\tilde{\lambda}$ until $\tilde{\lambda}=\infty$.
Therefore, we can deal with the range of the parameter
$0\leq\tilde{\lambda}<\infty$ by one potential function of
the swallow tail catastrophe.

When we fix  $\tilde{M}$ and   $\tilde{v}$, and change
 $\tilde{\lambda}$ from $0$ to $\infty$, the stable
branch of the NA black hole neither bifurcates nor merges
with another branch.  This means that the stability of a
NA black hole in the stable branch does not change.
Since Aichelburg and Bizon showed that the NA black hole
with $\tilde{\lambda}=\infty$ is stable\cite{AB}, we can
conclude that a NA black hole in the ``stable" branch is
always stable.

As for a non-singular monopole, as shown in some
references\cite{BFM,AB,Hollmann}, a cusp structure
appears in a \mbox{$\tilde{v}$-$M$} diagram when
$\tilde{\lambda}$ is small (Fig.~1).  Therefore, taking the
mass as a potential function and $\delta_H$ as a state
variable, we can understand the stability of the
non-singular monopole by fold catastrophe in the same way
as we treated neutral NA black holes\cite{MTTM,TMT}; the
lower branch is stable and the upper branch is
unstable\cite{Hollmann}.  We find that the
branch is really stable for the following reason.   The
branch changing
$\tilde{\lambda}$ with fixed $\tilde{v}$ and $\tilde{M}$
neither bifurcates nor merges with another branch.
Therefore, the stability of the branch does not change.
Since the monopole in the stable branch in the
\mbox{$\tilde{v}$-$M$} diagram is stable
when $\tilde{\lambda}$ is small\cite{Hollmann},
then the branch
is stable even when $\tilde{\lambda}$ is large.


\vskip 2 cm
\setcounter{equation}{0}
\section{NEW PROPERTIES OF A MONOPOLE BLACK HOLE}

In this section, we present new  properties of monopole
black holes which have not been discussed in \S. 2,
i.e., their thermodynamical properties and global
spacetime structure.

{\bf (1)} Thermodynamical properties:\\
First, we show the inverse temperature of the black holes in
Fig.~9.   For small values of $\tilde{\lambda}$, we find
the turning point $C$ in Fig.~9, which corresponds to the
cuspidal edge in a
\mbox{$\tilde{M}$-$\tilde{r}_H$} diagram in Fig.~3(b).   We
can understand this coincidence as follows:    Because the
black hole is non-rotating and its charge is fixed, the
first law of the black hole in the present system can be
written as\cite{HS}
\begin{equation}
  \delta M=T\delta S,
\end{equation}
where $S=\pi r_H^2$ is the entropy of the black hole.  From
this, we can write as
\begin{equation}
  \frac{dS}{dM}=\frac{1}{T},
\end{equation}
or
\begin{equation}
  \frac{dr_H}{dM}=\frac{1}{2\pi r_H T}.
\end{equation}
So, if  $1/T$ depends on $M$ as in Fig.~9,
that is, if  a turning point  $C$ exists, then a cusp must
appear in a \mbox{$M$-$r_H$} diagram.
Hence, the stability changes at the turning point $C$ on a
vertical tangent in the
\mbox{$M$-$1/T$} diagram, where
$d(1/T)/dM=\infty$ .  Thus,
we can also discuss the stability of a black hole by using
the \mbox{$M$-$1/T$} diagram.  As for this method, see
reference\cite{Katz}.

The specific heat of  monopole black holes in the stable
branch is always negative, while that in the unstable one
changes its sign a few times depending on the parameters
$\tilde{v}$ and $\tilde{\lambda}$.  This feature is similar
to that of neutral NA black holes.   When
\mbox{$\tilde{\lambda}<\tilde{\lambda}_{crit}(\tilde{v})$}
and $\tilde{v}$ is small enough, the sign of the specific
heat in the unstable branch changes at two points (one is
the turning point $C$ on the vertical tangent and the other
is another turning point $A$ on the horizontal tangent).
The specific heat between two turning points is
positive.  For example, for $\tilde{v}=0.05$ and
$\tilde{\lambda}=0.1$, we find the two turning points $C$
and $A$ at $\tilde{M}=0.32495$ and $\tilde{M}=0.32401$,
respectively.  Note that the  specific heat at $C$ vanishes
while that at $A$ diverges just like that of the RN black
hole with the critical charge.
 As
$\tilde{v}$ increases, the  unstable NA branch merges with
the RN branch before the specific heat turns to negative.
The sign of the specific heat changes only at the point
$C'$ as in Fig.9(a).  As $\tilde{v}$ increases further,
$\tilde{\lambda}_{crit}(\tilde{v})$ becomes smaller than
$\tilde{\lambda}$.  Then, the stability change does not
occur, and the specific heat is always negative.

{\bf (2)} Spacetime structure\\
Next, we consider the distribution of energy density around
a monopole to discuss the spacetime structure. The
energy density $\rho$ can be written as
\begin{equation}
  \rho=-{T^0}_0=\left( 1-\frac{2\tilde{m}}{\tilde{r}}
\right)
\rho_{\mbox{grad}}
  +\rho_{\mbox{pot}},  \label{den1}
\end{equation}
where the gradient term $\rho_{\mbox{grad}}$ and
the potential term$\rho_{\mbox{pot}}$  have forms
\begin{eqnarray}
  \rho_{\mbox{grad}} & = & e^2 v^4 \left\{
\frac{1}{\tilde{r}^2} \left(
\frac{dw}{d\tilde{r}}
  \right)^2 + \frac{1}{2} \left( \frac{dh}{d\tilde{r}}
\right)^2
  \right\}, \label{den2}
\\  \rho_{\mbox{pot}} & =  & \rho_{\mbox{pot}(F^2)}
+\rho_{\mbox{pot}(D\Phi^2)} +
  \rho_{\mbox{pot}(V)}, \label{den3}
\end{eqnarray}
with
\begin{eqnarray}
  \rho_{\mbox{pot}(F^2)} & = &
  e^2 v^4  \frac{(1-w^2)^2}{2\tilde{r}^4}, \label{den4}
\\
  \rho_{\mbox{pot}(D\Phi^2)} & = &
  e^2 v^4 \frac{h^2 w^2}{\tilde{r}^2}, \label{den5}
\\
  \rho_{\mbox{pot}(V)} & = &
  \frac{1}{4} \tilde{\lambda} e^2 v^4 (h^2-1)^2.
\label{den6}
\end{eqnarray}
$\rho_{\mbox{pot} (F^2)}$, $\rho_{\mbox{pot} (D\Phi^2)}$
and
$\rho_{\mbox{pot} (V)}$ come from the potential parts of
$-(1/4) F^2, ~ -(1/2) (D \Phi )^2$ and
$-(\lambda /4)(\Phi^2-v^2)^2$ in the matter Lagrangian,
respectively.

We show the  energy density distribution around the
non-singular monopole in Fig.10.    The energy density
is almost constant inside
the monopole core ($\tilde{r}
\mbox{\raisebox{-1.ex}{$\stackrel{\textstyle<}
{\textstyle\sim}$}}
1$), but it decays as
$1/\tilde{r}^4$ outside ($\tilde{r}
\mbox{\raisebox{-1.ex}{$\stackrel{\textstyle >}
{\textstyle\sim}$}} 1$). This indicates that
Eq.(\ref{simplifyed}) which Lee et al. used is a good
approximation.

This simplified model may tell us the global spacetime
structures of the gravitating monopole and its black
holes as follows:  To determine the mass function
$m(r)$ and the core radius $R$, we adopt the approximation of
(\ref{simplifyed}) and minimize $M$ .  Then, we
obtain\cite{LNW}:
\begin{equation}
  \frac{m(r)}{r}=\left\{ \begin{array}{ll}
       (M-M_{mon})/r+M_{mon}r^2/4R^3, &
                                         \mbox{($r<R$)} \\
       M/r-3M_{mon}R/4r^2, &
                                          \mbox{($r>R$)}
           \end{array} \right. \label{m/r}
\end{equation}
where $M_{mon} = 8\pi/3e^2R$
and $R= (2 e^2 \rho_0)^{-1/4}$.
Using the known relation
$M_{mon}=4\pi v/e$,
we can fix the relation between $\rho_0$
and $v$, i.e., $\rho_o=81 e^2v^4/32$.

{}From Fig.12, we can approximate a form of $\delta(r)$ to
be a step function:
\begin{equation}
  \delta=\left\{ \begin{array}{ll}
       \delta_0={\rm constant}, & \mbox{($r<R$)}
        \\
        \delta_{\infty}={\rm constant}, & \mbox{($r>R$)}
                 \end{array} \right. \label{delta}
\end{equation}
{}From our approximation (\ref{m/r}) and (\ref{delta}), we
regard the inside of the monopole core as the
Schwarzschild-de Sitter spacetime.
(de Sitter spacetime for ``monopole" solution,  $m(0)=0$)
and the outside as the Reissner-Nordstr\"{o}m spacetime.
Let us first discuss the case of the ``monopole"
($m(0)=0$).  The maximal value of
$\tilde{m}(\tilde{r})/\tilde{r}$ is
${\hat{M}_{mon}}^2/8\pi$ and
\mbox{$\tilde{m}(\tilde{R})/\tilde{R}= 3 {\hat{M}_{mon}}^2
/32\pi$}, where $\hat{M} \equiv eM/M_{Pl}$ and $\tilde{R}
\equiv evR$(see Fig.~13).   Since
$\tilde{m}_H/\tilde{r}_H=1/2$, if
\begin{equation}
  \frac{{\hat{M}_{mon}}^2}{8\pi}<\frac{1}{2}
          \Leftrightarrow
  \tilde{v} < \frac{1}{\sqrt{4\pi}},
\end{equation}
a horizon does not exist and the
surface ($R=\mbox{constant}$) is timelike.
The  Penrose diagram is shown in Fig.14(a).
When
\begin{equation}
  \frac{3{\hat{M}_{mon}}^2}{32\pi} < \frac{1}{2} \leq
         \frac{{\hat{M}_{mon}}^2}{8\pi}
                \Leftrightarrow
       \frac{1}{\sqrt{4\pi}} \leq \tilde{v} <
             \frac{1}{\sqrt{3\pi}}, \label{pen_reg2}
\end{equation}
horizons exist and $R=\mbox{constant}$ is
timelike(Fig.~14(b)), but when
\begin{equation}
  \frac{1}{2} \leq \frac{3{\hat{M}_{mon}}^2}{32\pi}
                \Leftrightarrow
   \tilde{v} \geq  \frac{1}{\sqrt{3\pi}}, \label{pen_reg3}
\end{equation}
horizons exist and $R=\mbox{constant}$ becomes spacelike.
See Fig.~14(c) for the Penrose diagram.
That is, the ``monopole" is no longer static and probably
the core itself expands exponentially.   This behavior may
suggest  a topological inflation\cite{top-inf}.    However,
following the discussion on stability of Cauchy
horizon\cite{Cauchy}, which exists in the present Penrose
diagram, the de Sitter phase in the black hole will
probably be unstable.
 The existence of a maximum value of $\tilde{v}$
($\tilde{v}_{\rm max}$) in a \mbox{$\tilde{v}$-$M$}
diagram (Fig.1) may also give some indication for the
topological inflation. Notice that the critical vacuum
expectation  value of the Higgs field is almost the same as
that in our simplified model.

Next we classify the spacetime of the black hole solutions
($m(0)\neq 0)$ in the same way  and describe the Penrose
diagrams(Fig.~15).   Using the simplified model, we find
the classification as follows:

\noindent(1)
\mbox{$\hat{M}<(3/2)\hat{M}_{mon}$} and
  \mbox{$\hat{M}<\sqrt{4\pi}$},

\noindent(2)
\mbox{$\hat{M}<(3/2)\hat{M}_{mon}$} and
   \mbox{$\sqrt{4\pi}
       <\hat{M}<4\pi/(3\hat{M}_{mon}) +
        (3/4)\hat{M}_{mon}$},

\noindent(3)
\mbox{$\hat{M}<(3/2)\hat{M}_{mon}$} and
  \mbox{$4\pi/(3\hat{M}_{mon})+(3/4)\hat{M}_{mon} <
   \hat{M} <
  (32\pi^{3/2})/(27\hat{M}_{mon}^2)+\hat{M}_{mon}$},

\noindent(4)
\mbox{$\hat{M}<(3/2)\hat{M}_{mon}$} and
   \mbox{$(32\pi^{3/2})/(27\hat{M}_{mon}^2)+\hat{M}_{mon}
   <\hat{M}$},

\noindent(5)
\mbox{$\hat{M}>(3/2)\hat{M}_{mon}$} and
       \mbox{$\hat{M}<4\pi/(3\hat{M}_{mon})
        +(3/4)\hat{M}_{mon}$},

\noindent(6)
\mbox{$\hat{M}>(3/2)\hat{M}_{mon}$} and
       \mbox{$4\pi/(3\hat{M}_{mon})
        +(3/4)\hat{M}_{mon}<\hat{M}$}.

\noindent We have required here the condition of no
 naked singularity, \mbox{$M>M_{mon}$}.     The
corresponding Penrose diagrams for the cases (1) $\sim$ (6)
are shown in Fig.~15.      If :

\noindent(7)\mbox{$\hat{M}<\hat{M}_{mon}$} and
  \mbox{$(3/2)\hat{M}_{mon}<\hat{M}$},

\noindent we have a naked singularity.

When we see the spacetime structures of
(\ref{pen_reg2}) and (\ref{pen_reg3}) for the ``monopole"
case, and (2), (3), (4), and (6) for the black hole case
from the outside of the event horizon, we cannot
distinguish them from the structure of the RN spacetime.
However, their global structures may be completely
different from the RN spacetime as shown in Fig.~14 and
Fig.~15.   In some spacetimes, a Cauchy horizon does not
appear.   The existence of the NA field changes the
global structure of the spacetime.

{\bf (3)} Physical understanding of the stabilities:\\
As discussed in Paper I, we may understand the physics of
the stabilities of NA black holes from a comparison of
energy distributions between stable and unstable branches.
For the stable and the unstable monopoles with the same
value of $\tilde{v}$(Fig.~11), we find that
$\rho_{\mbox{pot}(D\Phi^2)}$ is almost the same in the two
monopoles, while
\mbox{$(1-2\tilde{m}/\tilde{r})\rho_{\mbox{grad}}$} and
$\rho_{\mbox{pot}(F^2)}$ are larger  in the unstable one
than in the stable one.    We can regard
$\rho_{\mbox{pot}(D\Phi^2)}$ as a ``stabilizer" (see Paper
I).     From the Tolman-Oppenheimer-Volkoff equation of
hydrostatic equilibrium in the EYMH system\cite{KT}, the
term
\mbox{$(1-2\tilde{m}/\tilde{r})\rho_{\mbox{grad}}$} works
as an ``attractive force" and the term
$\rho_{\mbox{pot}(F^2)}$ works as a ``repulsive force".
Thus, we may say that the stronger  ``attractive force" by
the gradient term balances the stronger ``repulsive force"
by the magnetic field in an unstable monopole.  However,
since the ``stabilizer"($\rho_{\mbox{pot}(D\Phi^2)}$) of
the unstable monopole is almost the same as that of the
stable monopole, it cannot keep the balance of the
heavier burden(namely, the balance between the stronger
``attractive" and ``repulsive" force) stable.


\vskip 2 cm
\setcounter{equation}{0}
\section{SUMMARY AND CONCLUDING REMARKS}

We have re-investigated regular monopole and monopole black
hole solutions in the Einstein-Yang-Mills-Higgs system.
When we discussed neutral NA holes in our previous
papers\cite{MTTM,TMT}, we found that catastrophe theory is
applicable in analyzing the stability of NA
holes.   The properties of the monopole black hole are very
complicated and different from those of neutral holes in
many respects. However, we have found that catastrophe
theory is still applicable to the monopole black hole,
i.e., a swallow tail catastrophe can explain many features
of the monopole black hole very naturally.   It explains the
transition from the monopole black hole to the
Reissner-Nordstr\"{o}m black hole and its opposite
process. This means that we not only can understand its
stability, but also can get an insight into the structure of
the solution space.    Thus, catastrophe theory is not
only a powerful tool to study the stability of NA
black holes but also gives us a universal picture of
NA black holes.

The monopole black hole violates the weak no-hair
conjecture\cite{AB,grqc2016}, that is, two distinct stable
solutions can exist for a given mass.    But the monopole
black hole is not the first black hole violating the
weak no-hair conjecture.    The Skyrme black hole is also
stable against linear perturbation and the
non-singular Skyrmion may be stable even against  nonlinear
perturbation\cite{HSZ}.   However, the entropy of
the Skyrme black hole is always smaller than that of the
Schwarzschild black hole with the same mass.    Therefore,
the Skyrme hair would be lost in the formation process of
a black hole and it would become a Schwarzschild black
hole at last.     On the other hand, the monopole and the
monopole black holes are really classically stable because
they have the maximum entropy among black holes with the
same mass.  Such objects can be the final remnant in the
universe.  Even if we start with the RN black hole, it will
lose its mass energy via the Hawking  evaporation process
and a transition from the RN black hole to the monopole
black hole will occur at some critical point.  The
gravitating monopole may be found at last.    Thus the
monopole black hole may be the first real  example which
violates the no-hair conjecture and which can be formed in
the universe.

We have also discussed the thermal properties of monopole
black holes.    The sign of the specific heat of unstable
monopole black holes  changes one or two times depending on
the parameters.    On the other hand, the specific heat of
stable monopole black holes is always negative.     This
feature is the same as that found in neutral NA black holes.

We have shown  the energy density distributions of fields
for a gravitating monopole and its black holes and
discussed the global spacetime structure   by using a
simplified model.  If the vacuum expectation value of the
Higgs field $v$ is larger than a critical value ($\sim
0.370821 M_{Pl}$ for $\lambda=0.1$), the core of the
monopole cannot be static.  It may expand exponentially.
This may suggest a topological inflation, recently proposed
by
 Linde and Vilenkin\cite{top-inf}.


\vskip 2 cm
\setcounter{equation}{0}
\section{ACKNOWLEGEMENTS}
We would like to thank Gray W. Gibbons, Osamu Kaburaki,
Joseph Katz, Fjodor V. Kusmartsev, Takuya Maki, Ian Moss
and Eric Poisson for useful discussions.   This  work was
supported partially by the Grant-in-Aid for Scientific
Research  Fund of the Ministry of  Education, Science and
Culture   (No. 06302021 and No. 06640412),
 by the Grant-in-Aid for JSPS Fellows (053769),
and by the Waseda University Grant for Special Research
Projects.


\vspace{2cm}


\newpage
\baselineskip .65cm
\begin{flushleft}
{\bf Figure Captions}
\end{flushleft}
\baselineskip .65cm

\vskip 0.1cm
\noindent
\parbox[t]{3cm}{\bf Figure 1:\\~}\ \
\parbox[t]{12.5cm}{The $\tilde{v}-M$ diagram of a
gravitating  monopole for  $\tilde{\lambda}=0.1$ (the solid
line) and the extreme Reissner-Nordstr\"{o}m black hole (the
dotted line). }\\[1em]
\noindent
\parbox[t]{3cm}{\bf Figure 2:\\~}\ \
\parbox[t]{12.5cm}{The mass-horizon diagram
 of the Reissner-Nordstr\"{o}m black hole (the dotted
line) and of the monopole black hole (the solid line)
for  $\tilde{\lambda}=1$ and
$\tilde{v}=0.05.$}\\[1em]
\noindent
\parbox[t]{3cm}{\bf Figure 3:\\~\\~}\ \
\parbox[t]{12.5cm}{(a)The same diagram as Fig.2
for $\tilde{\lambda}=0.1$ and $\tilde{v}=0.05$.  (b)
The difference between the horizon radius of a monopole
black hole and that of a Reissner-Nordstr\"{o}m  black hole
is plotted near the bifurcation point $B$.  There exists a
small cusp structure.}\\[1em]
\parbox[t]{3cm}{\bf Figure 4:\\~}\ \
\parbox[t]{12.5cm}{The catastrophe set of
a swallow tail type.}\\[1em]
\noindent
\parbox[t]{3cm}{\bf Figure 5:\\~}\ \
\parbox[t]{12.5cm}{The behavior of a
potential function (entropy of the black hole) for
$\tilde{\lambda}>\tilde{\lambda}_{crit}$.The
maximum and minimum points correspond to
the stable and the unstable solutions,
respectively.  There are two solutions $R$ (the RN black
hole) and $N$ (the NA black hole).}\\[1em]
\noindent
\parbox[t]{3cm}{\bf Figure 6:\\~}\ \
\parbox[t]{12.5cm}{The same as Fig. 5
for
$\tilde{\lambda}<\tilde{\lambda}_{crit}$.
There are three solutions $R$ (the RN black
hole) and $N_1, N_2$ (the stable and unstable NA black
holes).}\\[1em]
\noindent
\parbox[t]{3cm}{\bf Figure 7:\\~}\ \
\parbox[t]{12.5cm}{The critical values of mass
$\tilde{M}_{\rm N-R}$, where the RN black hole and the NA
black hole merge, and
$\tilde{M}_{\rm N-N}$, where two NA solutions merge.
$\tilde{M}_{\rm N-R}$ is independent of
$\tilde{\lambda}$.}\\[1em]
\noindent
\parbox[t]{3cm}{\bf Figure 8:\\~}\ \
\parbox[t]{12.5cm}{The dependence of
$\tilde{M}_{\rm N-R}$ on $\tilde{v}$. The line represents
$\tilde{M}_{\rm N-R}=7.256
\tilde{v}^{1.745}+0.273.$}\\[1em]
\noindent
\parbox[t]{3cm}{\bf Figure 9:\\~}\ \
\parbox[t]{12.5cm}{The mass-inverse temperature diagrams
of the Reissner-Nordstr\"{o}m black hole (the dotted line)
and of the monopole black hole for
$\tilde{\lambda}=0.1$
and $\tilde{v}=0.03$ (the solid line),
$\tilde{v}=0.05$ (the dashed line),
$\tilde{v}=0.1$ (the dot  dashed line),
$\tilde{v}=0.15$ (the double-dot dashed line).
(b) The enlarged figure near the bifurcation point of
Fig.9(a) for $\tilde{\lambda}=0.1$ and $\tilde{v}=0.05$.
}\\[1em]
\noindent
\parbox[t]{3cm}{\bf Figure 10:\\~}\ \
\parbox[t]{12.5cm}{The energy density of the
monopole black hole for $\tilde{\lambda}=0.1$ and
$\tilde{v}=0.05$.  It is almost
constant inside the monopole core
$\tilde{r}\stackrel{<}{\sim}1$ but decays as
$1/\tilde{r}^4$ outside the core. }\\[1em]
\noindent
\parbox[t]{3cm}{\bf Figure 11:\\~}\ \
\parbox[t]{12.5cm}{(a) The potential parts of the energy
density ($\rho_{\mbox{pot}(F^2)}$ and
$\rho_{\mbox{pot}(D\phi^2)}$) of the stable monopole
(the solid line) and the unstable monopole (the dotted
line) for $\tilde{\lambda}=0.1$ and
$\tilde{v}=0.37083$.   We do not show the term
$\rho_{\mbox{pot}(V)}$ because it is small.
(b) The gradient part of the energy density
(\mbox{$(1-2\tilde{m}/\tilde{r})\rho_{\mbox{grad}}$}) of the
stable monopole (the solid line) and the unstable monopole
(the dotted line) for the same values of $\tilde{\lambda}$
and  $\tilde{v}$ as Fig. 11(a). }\\[1em]
\noindent
\parbox[t]{3cm}{\bf Figure 12:\\~}\ \
\parbox[t]{12.5cm}{The function $\delta(\tilde{r})$
for a gravitating  monopole for $\tilde{\lambda}=1$ and
$\tilde{v}=0.2$ (the solid line), $\tilde{v}=0.3$
(the dashed line),
$\tilde{v}=0.35$ (the dot  dashed
line), and $\tilde{v}=0.37$ (the dotted line).  As
$\tilde{v}$ increases, the approximation by a step
function may be justified. }\\[1em]
\noindent
\parbox[t]{3cm}{\bf Figure 13:\\~}\ \
\parbox[t]{12.5cm}{ The function form of
$\tilde{m}(\tilde{r})/\tilde{r}$.    The horizontal dot
dashed lines represent $\tilde{m}/\tilde{r}=1/2$ for
(i)$\tilde{v}<1/\sqrt{4\pi}$,
(ii) $1/\sqrt{4\pi} \leq \tilde{v}<1/\sqrt{3\pi}$, and
(iii) $\tilde{v} \leq \sqrt{3\pi}$. }\\[1em]
\noindent
\parbox[t]{3cm}{\bf Figure 14:\\~}\ \
\parbox[t]{12.5cm}{The Penrose diagrams of the gravitating
``monopole" when
(a) $\tilde{v}<1/\sqrt{4\pi}$,
(b) $1/\sqrt{4\pi} \leq\tilde{v}<1/\sqrt{3\pi}$, and
(c) $\tilde{v} \leq \sqrt{3\pi}$.   The dashed line
represents the surface of the ``monopole".   The shaded
and the unshaded regions are the de Sitter and the
Reissner-Nordstr\"{o}m spacetimes, respectively.
}\\[1em]
\noindent
\parbox[t]{3cm}{\bf Figure 15:\\~}\ \
\parbox[t]{12.5cm}{The Penrose diagrams of the monopole
black hole for (a) Case (1) and (5), (b) Case (2), (c) Case
(3), and (d) Case (4) and (6).    The shaded
and the unshaded regions are the Schwarzschild-de Sitter
and the Reissner-Nordstr\"{o}m spacetimes, respectively.
The double lines represent singularities. }\\[1em]
\noindent


\newpage
\vskip 2cm
\baselineskip .5cm
\begin{thebibliography}{99}
\bibitem{Dirac} P.A.M. Dirac, Proc. R. Soc. {\bf A133},
                60 (1931).
\bibitem{'tHooft-Polyakov} G. 't Hooft, Nucl. Phys.
                         {\bf B79}, 276 (1974);
                  A.M. Polyakov, JETP Lett., {\bf 20},
                  194 (1974).
\bibitem{top-inf} A. Linde, Phys. Lett. B {\bf 327}, 208
										(1994);  A. Linde and D. Linde,  Phys.  Rev.
          {\bf D50}, 2456 (1994);\\
       A. Vilenkin, Phys. Rev. Lett. {\bf 72}, 3137 (1994).
\bibitem{BFM}  P. Breitenlohner, P. Forg\'{a}cs and
              D. Maison, Nucl. Phys. {\bf B383}, 357(1992).
\bibitem{LNW} K.-Y. Lee, V.P. Nair and E. Weinberg, Phys.
              Rev. Lett. {\bf 68}, 1100(1992);
              Phys. Rev. {\bf D45}, 2751(1992);
             General Rel. Grav. {\bf 24}, 1203(1992).
\bibitem{Ortiz} M.E. Ortiz, Phys. Rev. {\bf D45},
                  R2586(1992).
\bibitem{AB}   P.C. Aichelburg and P. Bizon, Phys. Rev.
                  {\bf D48}, 607(1993).
\bibitem{TMT}  T. Torii, K. Maeda and T. Tachizawa,
               Preprint, WU-AP/39/94(1994),
               (gr-qc/9406013).
\bibitem{BM} R. Bartnik and J. McKinnon,
               Phys. Rev. Lett.
               {\bf 61}, 141(1988).
\bibitem{CBH}  M.S. Volkov and D.V. Galt'sov, Pis'ma
                Zh. Eksp. Teor. Fiz.
              {\bf 50}, 312(1989); Sov. J. Nucl. Phys.
              {\bf 51}, 74(1990)7;\\
                P. Bizon, Phys. Rev. Lett. {\bf 64},
                      2844(1990);\\
                H.P. K\"unzle and A.K. Masoud-ul-Alam, J.
                Math. Phys. {\bf 31}, 928(1990).
\bibitem{MTTM}  K. Maeda, T. Tachizawa, T. Torii and
                T. Maki,
                Phys. Rev. Lett. {\bf 72}, 450(1994).
\bibitem{notation} The vacuum expectation value $H_0$ which
                 Breitenlohner et al. used is related to
                 our $v$ by $H_0=\sqrt{4\pi}v$
\bibitem{LM} H.C. Luckock and I. Moss, Phys. Lett.
              {\bf B176}, 341(1986);
               H.C. Luckock, in
              {\it String Theory and Quantum Gravity} ed.
               by H.J.
               de Vega and N. Sanchez (World Scientific,
               1987) p. 455.
\bibitem{DHS} S. Droz, M. Heusler and N. Straumann, Phys.
                  Lett. {\bf B268}, 371(1991);\\
              P. Bizon and T. Chmaj, Phys. Lett.
              {\bf B297}, 55(1992).
\bibitem{TM}  T. Torii and K. Maeda, Phys. Rev. {\bf D48},
              1643(1993).
\bibitem{GMO} B.R. Greene, S.D. Mathur and C.M. O'Neill,
               Phys. Rev. {\bf D47}, 2242(1993).
\bibitem{Poston} T. Poston and I. Stewart, Catastrophe
                theory and its applications, Pitman
                publishing limited
\bibitem{Kusmartsev} F.V. Kusmartsev  Phys. Rep. {\bf 183},
                  1(1989)
\bibitem{Prasad_Sommerfield} M.K. Prasad and C.M.
                 Sommerfield, Phys. Rev. Lett. {\bf 35},
                 760(1975).
\bibitem{Davies} P.C.W. Davies, Proc. R. Soc. Lond.
                {\bf A353}, 499(1977).\
\bibitem{Katz} J. Katz, Mon. Not. R. astr. Soc. {\bf
                183}, 765(1978);
                     Mon. Not. R. astr. Soc. {\bf
                189}, 817(1979);\\
               O. Kaburaki, I. Okamoto and J. Katz,
               Phys.Rev. {\bf D47}, 2234(1993);\\
               J. Katz, I. Okamoto and O. Kaburaki, Class.
               Quantum Grav. {\bf 10}, 1323(1993)
\bibitem{grqc2016}  P. Bizon, preprint, gr-qc/9402016
                    (1994)
\bibitem{Hollmann} H. Hollmann, preprint,  MPI-PhT/94-31
              (gr-qc/9406018).
\bibitem{HS} M. Heusler and N. Straumann, Phys. Lett
              {\bf B315}, 55(1993)
\bibitem{KT} D. Kastor and J. Traschen, Pys. Rev.
             {\bf D46}, 5399(1992)
\bibitem{stable} We should note, however, that in general
    the minimal point does not always means a real
    stable configuration. We focus just on some specific
    mode and then  the other modes can be unstable.
    We use  ``stable"  in the sense that it is
    more stable for some mode than that in the other
    branch.
\bibitem{Wald} R. M. Wald, Phys. Rev {\bf D48},
              R3427(1993);
               V. Iyer and R. M. Wald, Phys. Rev {\bf D50}, (1994)
\bibitem{HSZ} M. Heusler, N. Straumann and Z.-h. Zhou,
              Helv. Phys. Acta {\bf 66}, 614(1993)
\bibitem{Cauchy} M. Simpson and R. Penrose, Int. J. Theor.
             Phys. {\bf 7}, 183(1973); \\
             J. M. McNamara, Proc. R. Soc. Lond.
             {\bf A358}, 499(1978);
              Proc. R. Soc. Lond. {\bf A364}, 121(1978); \\
             P. Brady and E. Poisson, Class. Quantum Grav.
             {\bf 9}, 121(1992)

\end{thebibliography}
\end{document}